\documentclass[linenumbers]{emulateapj}

\usepackage{lineno}

\newcommand\ho{\ifmmode {\rm HI} \else H{\small I} \fi}
\newcommand\hh{\ifmmode {\rm H_2} \else H$_2$ \fi}
\def\no{\ifmmode {N_{\rm HI}} \else $N_{\rm HI}$ \fi}
\def\nt{\ifmmode {N_{\rm H_2}} \else $N_{\rm HI}$ \fi}
\def\msun{\ifmmode {\rm M_{\odot}}\else $\rm M_{\odot}$\fi}
\def\mpc{\ifmmode {\rm M_{\odot} \ pc^{-2}} \else $\rm M_{\odot} \ pc^{-2}$ \fi}
\def\tra{\ifmmode  \text{HI-to-H}_2\else H{\small I}-to-H$_2$ \fi}
\def\aG{\ifmmode {\alpha G}\else $\alpha G$ \fi}
\def\iuv{\ifmmode {I_{\rm UV}}\else $I_{\rm UV}$ \fi}

\def\sg{\ifmmode \sigma_{g} \else $\sigma_{g}$ \fi}
\newcommand\hd{\ifmmode \textrm{HI-dust} \else H{\small I}-dust \fi}
\DeclareMathAlphabet{\pazocal}{OMS}{zplm}{m}{n}
\newcommand\ms{\ifmmode \pazocal{M}_s \else $\pazocal{M}_s$ \fi}
\newcommand\Nm{\ifmmode  N_{\rm M}  \else $N_{\rm M}$ \fi}

\usepackage{graphicx}
\usepackage{url}
\usepackage{verbatim}
\usepackage{amsmath}
\usepackage{color,soul}

\begin{document}

\title{The Molecular Cloud Lifecycle II: Formation and Destruction of Molecular Clouds Diagnosed via H$_2$ Fluorescent Emission}
\shorttitle{Observations of H$_2$ Emission: The Molecular Cloud Lifecycle  }

\author{Blakesley Burkhart \altaffilmark{1, 2}, Shmuel Bialy\altaffilmark{3}, Daniel Seifried\altaffilmark{4,5},  Stefanie Walch\altaffilmark{4,5}, Erika Hamden\altaffilmark{6}, \\  
Thomas J. Haworth\altaffilmark{7}, Keri Hoadley\altaffilmark{8}, Shuo Kong\altaffilmark{6}, 
Madisen Johnson\altaffilmark{1},
Sarah Jeffreson\altaffilmark{9},
Mark R. Krumholz\altaffilmark{10}, Min-Young Lee\altaffilmark{11, 12},
Amiel Sternberg\altaffilmark{13,2},
Neal J.\ Turner\altaffilmark{14}}

\altaffiltext{1}{Department of Physics and Astronomy, Rutgers University,  136 Frelinghuysen Rd, Piscataway, NJ 08854, USA}
\altaffiltext{2}{Center for Computational Astrophysics, Flatiron Institute, 162 Fifth Avenue, New York, NY 10010, USA}
\altaffiltext{3}{Physics Department, Technion - Israel Institute of Technology, Haifa, 31000, Israel}
\altaffiltext{4}{I. Physics Institute, University of Cologne, Z\"ulpicher Str. 77, 50937 Cologne, Germany}
\altaffiltext{5}{Center for Data and Simulation Science, www.cds.uni-koeln.de, Cologne, Germany}
\altaffiltext{6}{Steward Observatory, University of Arizona, Tucson, AZ 85719, USA}
\altaffiltext{7}{Astronomy Unit, School of Physics and Astronomy, Queen Mary University of London, London E1 4NS, UK}
\altaffiltext{8}{Department of Physics \& Astronomy, University of Iowa, Van Allen Hall, Iowa City, IA 52242, USA}
\altaffiltext{9}{Harvard-Smithsonian Center for Astrophysics, 60 Garden Street, Cambridge, Massachusetts, USA}
\altaffiltext{10}{Research School of Astronomy and Astrophysics, Australian National University, Canberra ACT 2600 Australia}
\altaffiltext{11}{Korea Astronomy and Space Science Institute, 776 Daedeok-daero, Daejeon 34055, Republic of Korea}
\altaffiltext{12}{Department of Astronomy and Space Science, University of Science and Technology, 217 Gajeong-ro, Daejeon 34113, Republic of Korea
}
\altaffiltext{13}{Tel Aviv University, P.O. Box 39040, Tel Aviv 6997801, Israel}
\altaffiltext{14}{Jet Propulsion Laboratory, California Institute of Technology, Pasadena, CA 91109, USA}

\shortauthors{Burkhart et al. }

\begin{abstract}
Molecular hydrogen (H$_2$) formation and dissociation are key processes that drive the gas lifecycle in galaxies. 
Using the SImulating the LifeCycle of Molecular Clouds (SILCC) zoom-in simulation suite, we explore the utility of future observations of  H$_2$ dissociation and formation for tracking the lifecycle of molecular clouds.  The simulations used in this work include non-equilibrium H$_2$ formation, stellar radiation, sink particles, and turbulence.   We find that,
 at early times in the cloud evolution,  H$_2$ formation rapidly outpaces dissociation and molecular clouds build their mass from the atomic reservoir in their environment. Rapid H$_2$ formation is also associated with a higher early star formation rate. For the clouds studied here, H$_2$ is strongly out of chemical equilibrium during the early stages of cloud formation but settles into a bursty chemical steady-state about 2 Myrs after the first stars form. At the latest stage of cloud evolution, dissociation outweighs formation and the clouds enter a dispersal phase.
We discuss how theories for the molecular cloud lifecycle and the star formation efficiency may be distinguished with observational measurements of H$_2$ fluorescence with a space-based high-resolution FUV spectrometer, such as the proposed Hyperion and Eos NASA Explorer missions. Such missions would enable measurements of the H$_2$ dissociation and formation rates, which we demonstrate can be connected to different phases in a molecular cloud's star-forming life, including cloud building, rapidly star-forming, H$_2$ chemical equilibrium, and cloud destruction.

\end{abstract}

\maketitle

\section{Introduction}
Molecular hydrogen (H$_2$) is the most abundant molecule in the universe and is a critical component in the lifecycle of baryons throughout cosmic time.  In particular, H$_2$ is an important cooling channel in the early universe \citep{2013RPPh...76k2901B, 2019ffbh.book...67K, Bialy2019ApJ...881..160B, doi:10.1146/annurev-astro-082812-141034, 2022ApJ...927L..12N} and cold clouds of molecular H$_2$ are the medium through which all known star and planet formation occurs \citep{Chevance2022arXiv220309570C}.  In galaxies, indirect measurements of the mass of H$_2$ are strongly correlated with the star formation rate and used to estimate the star formation depletion times and efficiencies \citep{2012ApJ...745..190L,2014PhR...539...49K}.
The presence of H$_2$ molecules is fundamental to the formation of other heavy molecules such as CO, OH, HCN, and H$_2$O that serve as efficient coolants of dense gas \citep[e.g.,][]{1973ApJ...185..505H, 1995ApJS...99..565S, 2013RvMP...85.1021T,2015MNRAS.450.4424B}. 

On galaxy scales, observations reveal that the star-formation rate (SFR) surface density ($\Sigma_{\rm SFR}$) is strongly correlated with the H$_2$ mass surface density ($\Sigma_{\rm H_2}$) \citep{Bigiel2008AJ....136.2846B, 2010MNRAS.407.2091G,2011AJ....142...37S,2013ApJ...768...74T,2016NatSR...626896A,2022arXiv220402511G}.
However, it has not been observationally demonstrated if star formation is first triggered by gas cooling via the process of H$_2$ formation or if the gas is already self-gravitating before efficient cooling takes place. 
Simulations show that the two processes most likely happen hand-in-hand at solar metallicity in similar physical environments of high density, UV-shielded gas \citep{2010ApJ...709..308M, 2021MNRAS.505.1678J}. However, other simulations suggest that the atomic hydrogen phase is already globally collapsing \citep{2020ApJ...903...46C}. Additional theoretical work has demonstrated that H$_2$ is not needed for star formation in low metallicity environments \citep{2012MNRAS.421....9G,krumholz2013MNRAS.436.2747K}. Further observations of molecular gas at the H\,\textsc{i} to H$_2$ phase transition are required in order to clarify the full picture. 

Despite the critical importance of H$_2$ physics to star and planet formation, the relationship between \textit{the formation and destruction rates of molecular hydrogen and the star formation rate} is not well constrained \citep{Dalgarno1976MoleculeFI,Cazaux2002MolecularHF,Jura1974FormationAD,2017MolAs...9....1W}. 
Piecing together how H$_2$ affects star formation in galaxies is complicated by the fact that most H$_2$ is difficult to observe directly. The lowest-lying excited state that is capable of radiating, the (v,J) = (0,2) state, is at T=511K while the bulk of the molecular material is cold with temperatures of T$\approx10K$. Thus, warm H$_2$ is readily observed in the IR (e.g. by JWST  at z=0) in places where the gas is shock-heated and can be collisionally excited. However, to observe the cold H$_2$ gas, one must typically use less abundant molecular tracers that are easily accessible and bright at these lower temperatures. For example, the lowest-lying level of CO, (v,J) = (0,1), is at 5.5 K. Hence, CO emission is commonly converted to an approximate H$_2$ mass using a scaling factor (e.g., X$_{CO}$, \citealt{2013ARA&A..51..207B,Sandstrom2013ApJ...777....5S,2014ApJ...785L...1C,2022ApJ...931...28H,2022arXiv220402511G}). 

An underutilized method for directly observing the lifecycle of H$_2$ at cloud boundaries does exist:  H$_2$ does have significant sets of lines that can be produced via \textit{fluorescence} \citep{1990Martin,1989Sternberg}.  H$_2$ in the ground state absorbs far-ultraviolet (FUV) photons of wavelength $\lambda >$ 912 {\AA}, which excite the molecules into electronically excited states. In about ~15\% of cases, this results in the dissociation of the molecule.  ~85\% of cases produce a de-excitation cascade to vibrationally excited states resulting in FUV emission lines at wavelengths between 912 and 1700 {\AA} via transitions from excited electronic states ($B^1\Sigma^+_u$, $C^1\Pi^u$) to the ground state. The transitions within the vibrational rotational energy levels of the ground electronic state result in quadrupole transition lines at the near- and mid-infrared wavelengths \citep{1987ApJ...322..412B,1989Sternbergb}. The FUV lines, in particular, form a rich spectrum in the Lyman and Werner bands (photons with energies of 11.2 to 13.6 eV).  While the most common excitation mechanism is an external UV source, cosmic rays and x-rays, along with secondary electrons, can also produce the emission \citep{1988Sternberg,1997ApJ...481..282T,2020CmPhy...3...32B,2022A&A...664A.150G}.

The challenge for observations of this emission is that they must generally be done from space or a stratospheric balloon because the FUV photons are blocked by the atmosphere and the IR photons are masked by bright foregrounds.
FUV fluorescent H$_2$ emission lines have been observed in nearby molecular clouds, including Taurus, Ophiuchus, Orion, as well as several superbubbles and emission nebulae \citep{2004ApJ...616..257F,2005ApJ...629L..97F,2006ApJ...644L.181L,2006ApJ...644L.185R,2015MNRAS.449..605L,2011ApJ...738...91J,2015ApJ...807...68J}.
 An all-sky low spatial and spectral resolution map of FUV fluorescent H$_2$ emission lines was published by \citet{jo2017ApJS..231...21J} using the FIMS/Spear dataset \citep{2006ApJ...644L.159E,2011ApJS..196...15S}. 
The observations of H$_2$ FUV lines thus far are at limited spatial and spectral resolution. This situation may soon change with new FUV telescopes, such as the proposed Hyperion and Eos mission concepts \citep{2022JATIS...8d4008H}, which would use long-slit high-resolution spectroscopy to observe emission from H$_2$.
Such a mission would allow us to directly probe the H$_2$-H\,\textsc{i} conversion process (i.e. H$_2$ dissociation) and indirect probes of H$_2$ formation. While IR lines are observable in these clouds, they don't tell a complete picture. The FUV fluorescent lines give a direct tracer of the incident energy field since they are directly excited. The IR ro-vibrational lines are significantly weaker than the UV lines, making them challenging to observe. In addition, IR ro-vibrational lines often require special conditions to arise; In most situations where IR-H$_2$ lines are observed, the role of UV-pumping is just one of several possible pumping mechanisms, which additionally makes interpretation of the conditions creating IR-fluorescence challenging \citep{2022JATIS...8d4008H}. But a combined observing campaign measuring IR and UV H$_2$ fluorescence would provide a powerful look into H$_2$ clouds- providing boundary conditions, radiation fields, and information on the state of the gas deeper within the cloud.

In anticipation of such a telescope mission, we present a series of papers exploring star formation and  H\,\textsc{i}-H$_2$ transition science. Specifically, in this paper and in the companion paper (Bialy et al. 2023, hereafter, paper I) we explore the utility of FUV space observations for directly determining the formation and dissociation rates of H$_2$. 
Here we use the SILCC ( SImulating the LifeCycle of molecular Clouds) Zoom simulations to determine how the formation and dissociation rates of H$_2$ are connected to the underlying star formation rate (SFR) and lifecycle of H$_2$ in the cloud.  The SILCC-Zoom simulations include a non-equilibrium on-the-fly chemistry
network describing the formation of H$_2$ and CO. The simulations thus provide detailed thermodynamic and chemical conditions for self-consistent modeling of the
formation and destruction rates of H$_2$ out of the diffuse ISM and are an ideal dataset for studying the observational signatures and implications of FUV tracers of H$_2$. These rates, along with their connection to the star formation rate, are the critical determinants of molecular cloud lifetimes and set the initial conditions for complex chemistry in protoplanetary disks. 
Furthermore, the rates measure the sources and sinks of molecular mass in galaxies.  In a companion paper (paper I), we explore the H$_2$ formation and destruction spatial distributions in evolving clouds and how these may be obtained using FUV and 21 cm observations.

This paper is organized as follows: In Section \ref{sec:num} we describe the SILCC-Zoom simulations studied here to test the utility of the FUV to measure the lifecycle of H$_2$. 
  In Section  \ref{sec:sfr}
we present our results on  the connection of H$_2$ formation and dissociation rate with the star formation rate in the numerical simulations. 
In Section \ref{sec:dis} we discuss our results in the context of future and past studies and draw our conclusions in Section \ref{sec:con}.

\section{Numerical Simulations}
\label{sec:num}
\subsection{Simulation Overview}
The simulations used for this study are the SILCC-Zoom simulations \citep{Seifried2017MNRAS.472.4797S}, whose initial conditions are the galactic-scale SILCC simulations first presented in detail in \citet{Walch2015MNRAS.454..238W} and \citet{Girichidis2016MNRAS.456.3432G}. We study two specific runs with stellar feedback, which are described in detail in  \citet{haid2019MNRAS.482.4062H} and \citet{Seifried_2020} and are denoted as MC1-HD-FB and MC2-HD-FB (hereafter, MC1 and MC2). In the following, we briefly describe the numerical methods.

These simulations use the adaptive mesh refinement (AMR), finite-volume
code FLASH version 4 \citep{2000ApJS..131..273F,2008PhST..132a4046D}  to solve the magnetohydrodynamic equations.  The original stratified box SILCC runs, which are the initial conditions for the zoom-ins, have a  base-grid resolution of 3.9 pc and the zoom-in simulations used here resolve individual molecular clouds with up to a spatial resolution of 0.12~pc.
Both self-gravity of the gas \citep[][]{2018MNRAS.475.3393W} as well as a background potential for the stellar component of the
Galactic disk (modelled with an isothermal sheet with $\Sigma_{\rm star}=30$  M$_{\sun} \rm{pc}^{-2}$
and a scale height of 100 pc) are included.
Before starting to zoom in, turbulence is generated in the simulations by inserting supernovae (SNe) at random positions, with each SN injecting
$10^{51}$~erg of thermal energy or heat. The SN injection rate is set by the SFR following the Kennicutt-Schmidt relation \citep{1998ApJ...498..541K}, $\Sigma_{SFR} \propto \Sigma_{gas}^{1.4}$ and assuming a \citet{2001ApJ...554.1274C} initial mass function. When subsequently simulating the evolution of the molecular cloud in the zoom-in simulation, we turn off this SN feedback.

The chemistry of the ISM is modeled using a network for hydrogen and carbon chemistry \citep{1997ApJ...482..796N, 2007ApJS..169..239G,2010MNRAS.404....2G, 2012MNRAS.421..116G}. 
The network tracks the evolution of the chemical abundances of e$^-$, O, H$^+$, H, H$_2$, C$^+$, and CO. 
The chemical network also follows the thermal evolution of
the gas, i.e., the heating and cooling processes using the
chemical abundances provided by the network \citep{2010MNRAS.404....2G,2015MNRAS.454..238W}. For the heating via the
photoelectric effect as well as photodissociation reactions, the simulations include a uniform interstellar radiation field (ISRF) with
a strength of $G_{0}$ = 1.7 \citep{1978ApJS...36..595D} in units of the Habing field, which resembles the typical intensity of the FUV interstellar radiation field.  We will use the notation of G$_0$ as our standard unit for radiant intensity throughout this paper. The ISRF is attenuated due to shielding by the surrounding gas and dust. The necessary column densities of H$_2$, CO,
and the total gas are calculated via the TreeCol algorithm \citep{2012MNRAS.420..745C,2018MNRAS.475.3393W}. 

These SILCC-Zoom runs use a feedback prescription described in \citet{haid2019MNRAS.482.4062H}.
Sink particles are used to
model the formation of stars or star clusters and their
subsequent radiative stellar feedback. 
A sink particle can only be formed in a computational cell if the cell lives on the highest refinement level in the AMR grid. The accretion radius is set to 0.31~pc. We further demand that the gas is Jeans unstable, is in a converging flow, and represents a local gravitational potential minimum. Sink particles accrete gas and set the instantaneous star formation rate. A fraction of the accreted gas is turned into massive stars by means of the star cluster sub-grid model.
The star formation rate used throughout the paper is SFR = $\Delta M_{sink}/\Delta t$,  where $\Delta t$ is the time between the current and previous snapshot, which is about 3.3~kyr.

Each massive star represented in the sink follows its mass-dependent stellar evolutionary track, in which the amount of photoionizing radiation released by
each star is accounted for. The radiative feedback is
treated with a backward ray-tracing algorithm, TreeRay \citep{2021MNRAS.505.3730W}, and the radiative
transport equation is solved for hydrogen-ionizing EUV radiation assuming the On-the-Spot approximation with a temperature-dependent recombination coefficient. The resulting number of hydrogen-ionizing photons \citep{2018MNRAS.478.4799H}
and the associated heating rate are processed within the
chemical network. 
In addition to the ISRF, cosmic rays (CRs) and stellar EUV and FUV radiation from massive stars can also dissociate molecular hydrogen and are included in our dissociation rate calculations. The cosmic ray ionization rate with respect to atomic hydrogen is set to
a constant value of $3\times 10^{-17}$~s$^{-1}$ in the entire simulation domain.  
For more details on the simulations see \citet{haid2019MNRAS.482.4062H} and \citet{Seifried_2020}.

\subsection{H$_2$ formation and dissociation rates }

 H$_2$ formation occurs primarily from catalytic reactions on dust grains, a process first proposed by \citet{1948HarMo...7...73V}, with rates later calculated by \citet{1960MNRAS.121..238M}.  There are other channels to form H$_2$ in pure gas-phase through the H$^-$ anion \citep{1961Obs....81..240M,1973ApJ...181...95D,Sternberg_2021}, but these reactions are slower in most local ISM conditions than those via dust surface formation \citep{2003ApJ...584..331G}.  Spontaneous radiative dissociation of H$_2$ consists of breaking the covalent bond between the two hydrogen atoms, usually via excitation into the Lyman and Werner bands by an external UV source.

For the H$_2$ formation rates coefficient (R) and H$_2$ photodissociation rate (D) we follow the assumptions made in the SILCC simulation, and adopt a H$_2$ formation rate coefficient \citep[see][]{1989ApJ...342..306H,2010MNRAS.404....2G}:
\begin{align}
    R(T, T_d) &= 3 \times 10^{-17} T_2^{1/2} f_a(T_d) S(T,T_d) \ {\rm cm^3 \ s^{-1}} \\ \nonumber
    f_a(T_d) &\equiv \left(1 + 10^4 \mathrm{e}^{-600 {\rm K} / T_d}\right)^{-1} \\ \nonumber
    S(T,T_d) &\equiv \left(1 + 0.4 (T_2+T_{d,2})^{0.5} +
                   0.2 T_2 + 0.08 T_2^2 \right)^{-1} \ ,
\end{align}
where $T$ and $T_d$ are the gas and dust temperatures,  S is the sticking coefficient (the probability that a hydrogen atom striking the grain will stick to the surface), and f$_a$ is the fraction of adsorbed hydrogen atoms that actually form H$_2$, rather than simply escaping back into the gas phase. Here, $T_2$ denotes the temperature normalized to 100 K.

For the local H$_2$ dissociation rate per molecule, we include three different processes. The first is the dissociation by the ISRF given by
\begin{align}
    D_\mathrm{ISRF} = 3.3 \times 10^{-11} G_0 f_{\rm H_2 shield} \mathrm{e}^{-3.5 A_{\rm V, eff}} \ {\rm \ s^{-1}}  ,
\end{align}
where $G_0$ = 1.7 indicates the strength of the incident, unattenuated radiation field upon the cloud, $f_{\rm H_2 shield}$  and $\mathrm{e}^{-3.5 A_{\rm V, eff}}$ are the attenuation of the dissociation rate due to H$_2$ self-shielding and dust absorption, respectively, calculated via the TreeCol algorithm. The attenuation factors depend on the density structure of the cloud, as well as on the abundance of H$_2$ in each cell. The second is given by the cosmic ray ionization rate: 
\begin{align}
    D_\mathrm{CR} = 6 \times 10^{-17} \ {\rm \ s^{-1}}  .
\end{align}
The third process is caused by  UV radiation from stars and is given by \citep{Tielens2005}:
\begin{align}
D_\mathrm{stellar} = \frac{F_\mathrm{>13.6 eV}}{\mathrm{cm^{-2} \; s^{-1}}} \times \frac{\sigma_{\mathrm{H_2}}}{{\mathrm{cm}^{2}}} \times \left(\frac{\mathrm{13.6 eV}}{ E_\gamma }\right)^{3} \ {\rm \ s^{-1}} .
\end{align}
Here, $F_\mathrm{>13.6 eV}$ is the local flux density
of all photons with an energy above 13.6~eV calculated with our radiation transport algorithm TreeRay \citep{2021MNRAS.505.3730W}, and $ E_\gamma $ the average photon energy of these photons. We use a cross-section for the dissociation, $\sigma_{\mathrm{H_2}}$, of \mbox{6.3 $\times$ 10$^{-18}$ cm$^{-2}$} at 13.6~eV, decreasing with the photon energy to the power of -3.
The sum of the three dissociation rates gives the total dissociation rate
\begin{align}
    D = D_\mathrm{ISRF} + D_\mathrm{CR} + D_\mathrm{stellar} \ .
\end{align}

The formation and dissociation rate densities per unit volume are therefore: 
\begin{align}
j_F = n({\rm H})n_{H,tot} R \\
j_D = n({\rm H_2}) D
\end{align}
For every time snapshot in the simulation, the outputs provide the atomic hydrogen number density $n({\rm H})$, the ionized hydrogen number density $n({\rm H^+})$, and molecular hydrogen number density $n({\rm H_2})$, where the total hydrogen density is $n_{H,tot}= n({\rm H}) + 2n({\rm H_2}) + n({\rm H^+})$. Other outputs include $T$, $T_d$, $f_{\rm H_2 shield}$ and  $A_{\rm V, eff}$ for each cell, allowing us to calculate $j_F$ and $j_D$ on a cell-by-cell basis.
The ratio $j_F/j_D$ is a measure of how close/far each region is to steady-state.
In steady state, $j_F/j_D=1$, and in this limit $n({\rm H_2})/n({\rm H}) = Rn_{H,tot}/D$.
On the other hand, when the system is out of steady state, then $j_F \neq j_D$. If there is a net H$_2$ formation, $j_F/j_D>1$, and if there is net destruction $j_F/j_D<1$. Paper I (Bialy et al. in prep.) explores under what conditions a steady state can be assumed in these simulations and where the H$_2$ is strongly out of equilibrium.

\section{The Molecular cloud lifecycle and star formation }
\label{sec:sfr}

\subsection{Sources and Sinks of H$_2$}

For the remainder of this paper, we explore the time evolution of H$_2$ formation and dissociation in these simulations averaged over the entire box (i.e., the volume integrals of $j_F$ and $j_D$).  We convert these rates to total mass per time forming or dissociating H$_2$ and denote these single rate values as  $J_F$ and  $J_D$:
\begin{align}
J_F = m_{H_2}\int j_F \mathrm{d}V = m_{H_2}\int R n(H) n_{H,tot} \mathrm{d}V \label{eq:form}\\
J_D = m_{H_2}\int j_D \mathrm{d}V = m_{H_2}\int D n(\rm H_2) \mathrm{d}V 
\label{eq:dis}
\end{align}
We can also track, over the cloud lifetime, the mass budget of H$_2$ and study the cloud growth and dissociation in relation to the overall star formation activity. Ultimately, our goal is to see how the H$_2$ dynamics relate to the cloud lifetime and evolution of both the cloud and the stars formed within.

\begin{figure*}
    \centering
    \includegraphics[width=1.0\textwidth]{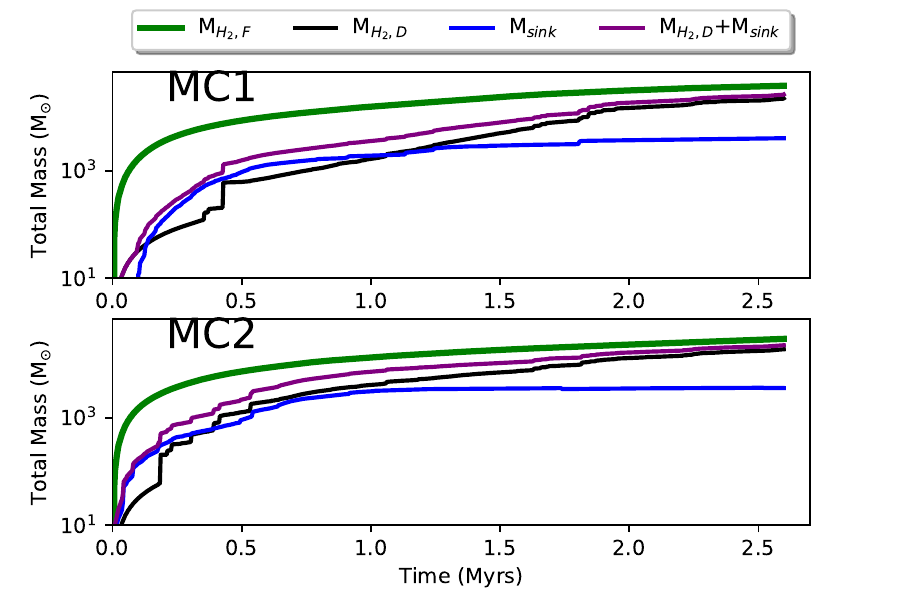}
   \caption{ Total mass of H$_2$ formed (green line), total mass of H$_2$ dissociated (black line), total sink mass formed (blue line), and total sinks of  molecular gas (purple line) for MC1 (top panel) and MC2 (bottom panel) vs. time.
}
    \label{fig:int} 
\end{figure*}

First, we investigate how the total mass of molecular hydrogen and stellar sink mass evolves with time. Throughout the paper, $t$ = 0 refers to the time when the first sink particle is formed, which happens about 1.5~Myr after we start to zoom in on the individual clouds in the galactic-scale SILCC simulation.
Figure \ref{fig:int} shows the total mass of H$_2$ formed (green line), total mass of H$_2$ dissociated (black line), total sink mass formed (blue line), and total sinks of  molecular gas (purple line) for MC1 (top panel) and MC2 (bottom panel) vs. time. Note that the values shown refer to times after $t$ = 0, i.e. H$_2$ formation/destruction before that time is not accounted for in this plot and thus, the values show the total change relative to the state at $t$ = 0. 

Total sinks of molecular gas refers to the contributions of H$_2$ dissociation and sink formation. For both clouds in the first several Myrs, molecular formation dominates over star formation and dissociation. The total mass dissociated or input into newly formed stars approaches the total mass of H$_2$ formed at around 2 Myrs into the starforming life of the clouds.  From Figure~\ref{fig:int} it is also clear that,   averaged over a long time, at the latest stage, dissociation outweighs formation as the violet line approaches the green line and the slope is steeper. This signals that the clouds are going into the dispersal phase.
The growth in dissociation is due to additional UV radiation from forming massive stars. Once stars form the stellar contribution to the dissociating radiation outpaces the contribution from ISFR and CRs. The total mass of H$_2$ dissociated outpaces the mass input into stars at around 1Myr for each cloud.   The total formation rate of MC1 is 1.3 times more than that of MC2. At the end of the run,  MC1 has  60\% more mass in stars than MC2, and  the stellar UV luminosity is about 60\% higher, as shown in \citet{2022arXiv220402511G}.

\begin{figure*}
    \centering
    \includegraphics[width=1.0\textwidth]{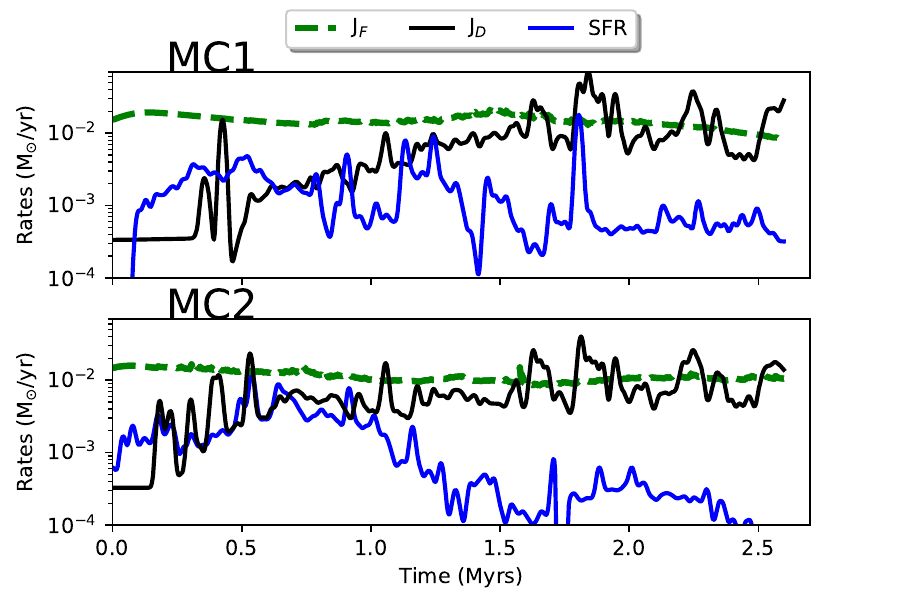}
   \caption{The time evolution of the total H$_2$ formation rate in the cloud (green dashed line), H$_2$ dissociation (black line), and star formation rates (blue line).  SFR and  H$_2$ dissociation rates have been smoothed with a spline function. Top row shows results for MC1 and bottom row shows results for MC2.}
    \label{fig:mc1} 
\end{figure*}

Figure \ref{fig:mc1} shows the time evolution of H$_2$ formation and dissociation rates integrated over the entire cloud volume (i.e., Equations \ref{eq:form} and \ref{eq:dis}) and the total SFR in MC1 and MC2.  The H$_2$ formation rate (green dashed line) is significantly more smoothly varying in time than the SFR (blue line) and the H$_2$ dissociation rate (black line). 
We note that the green dashed line shows the rate at which the molecular cloud is building mass while the two solid lines represent processes that remove mass from the molecular gas. 
During the early phases of cloud evolution (less than 1 Myr), active star formation leads to a corresponding increase in photodissociation, as the UV from newly formed stars impacts the cloud. Enhancements in the H$_2$ dissociation rate and the SFR are bursty. The dissociation rate and SFR are slightly correlated for MC1 (correlation coefficient is 0.25) but not correlated in the case of MC2. At early times in both zoom-in clouds, the H$_2$ formation rate significantly outpaces the H$_2$ dissociation rate.   The H$_2$ formation rate stays relatively constant throughout the cloud lifetime (i.e., see also Figure \ref{fig:int}), however MC1 shows a slight continuous decrease in  $J_F
$  past 1.8 Myr.  This suggests the cloud has entered a quenching phase in which it is unable to build new H$_2$ from the outside. 
 Regarding MC2, it is  continuously accreting from the cold neutral atomic hydrogen (CNM) reservoir outside of the cloud, forming H$_2$ in a slightly more rapid manner than MC1 \citep[see][]{2022arXiv220402511G}. This reservoir of cold atomic gas keeps the H$_2$ formation rate nearly constant in MC2.
 At later times, ongoing star formation creates a situation  where H$_2$ dissociation begins to dominate over H$_2$ formation, albeit in a bursty nature.  
Both clouds undergo a significant star formation event around 1.8 Myrs after which the SFR jumps several orders of magnitude. This leads to a jump in the amount of dissociating UV photons present in and around the clouds.
Overall, this matches the result that the time-integrated amount of destroyed H$_2$ (violet line in Figure~\ref{fig:int}) catches up with the formed H$_2$ (green line) at later times.

\begin{figure*}
    \centering
    \includegraphics[width=1.0\textwidth]{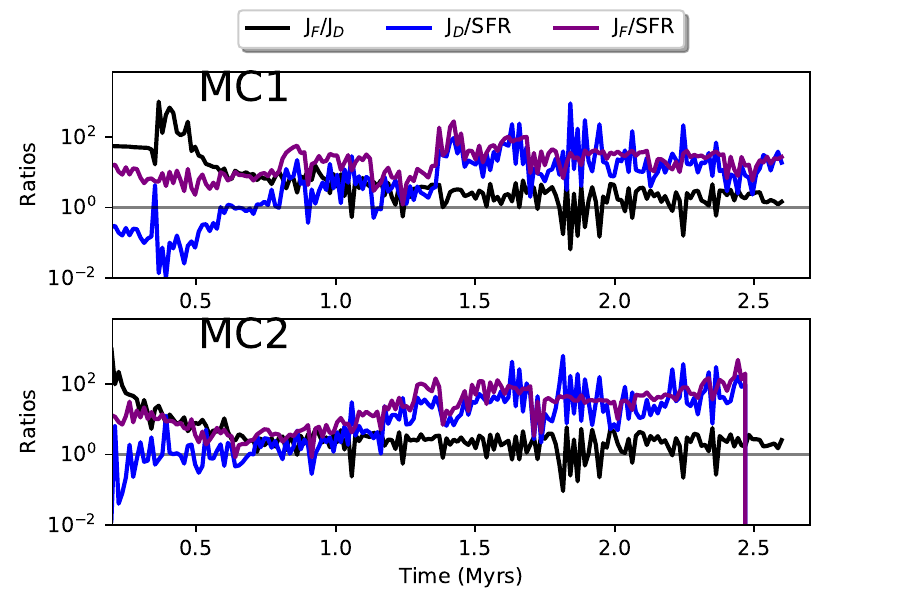}
   \caption{ Ratios of  $J_F$ to $J_D$ (black line), $J_F$ to SFR (purple line), and  $J_D$ to SFR (blue line) for MC1 (top panel) and MC2 (bottom panel) vs. time.
}
    \label{fig:ratios} 
\end{figure*}

To further understand the interplay of sources and sinks of molecular gas, we plot the ratios of $J_F $ to $J_D $, $J_F $ to SFR, and  $J_D $ to SFR in Figure \ref{fig:ratios}.  As seen in the previous two Figures, for both clouds,   $J_F $ dominates over the $J_D $, particularly at early times, and the clouds are strongly out of H\,\textsc{i} to H$_2$ transition equilibrium. Furthermore, H$_2$ formation is continuously higher than star formation and thus the cloud is building itself faster than stars can form. Such a conclusion points to the fact that the integrated star formation efficiency (the total mass of stars relative to the \textit{initial} H$_2$ gas mass) is only a useful theoretical construct for a closed system \citep[in agreement with][]{2023arXiv230110251J}. In fact, in the first 0.5 Myr of the cloud evolution, the $J_F $ dominates over $J_D $ by well over an order of magnitude.    $J_D$/ SFR  represents the total ``sinks" of H$_2$ gas (i.e., being removed from the cloud by going into stars or by being dissociated back into atomic H\,\textsc{i}). Star formation dominates as the sink of H$_2$ gas only in the very early phases of the cloud's lifetime, where H$_2$ photodissociation is generated mainly by the ISRF and the SFR is high (i.e., J$_D$/SFR $<1$). 

Our results demonstrate that the assumption of H$_2$ equilibrium in molecular clouds is incorrect at early stages (first few Myrs) in the cloud lifetimes.  This is in agreement with \citet{2021ApJ...920...44H}, \citet{2022MNRAS.512.4765S}, and \citet{2022arXiv220606393E}, who also found that equilibrium chemistry assumptions (very often employed in galaxy/cosmological simulations to track chemistry) at early stages in the cloud evolution are highly questionable. However, after about 2 Myrs into the cloud's star-forming evolution, the clouds studied here tend towards equilibrium, with J$_F$/J$_D \approx 1$.

Our analysis is primarily sensitive to photo-dissociation of H$_2$ (ISRF plus UV from stars). We included the contribution of H$_2$ destruction via cosmic rays; however, we find that the CR dissociation rate is several orders of magnitude lower than that of the UV field, when averaged over the entire cloud.  CR dissociation can, however, become important in very dense regions that are shielded from the ISRF UV and stellar UV fields \citep{2020CmPhy...3...32B}.

\subsection{Diagnosing Molecular Cloud Evolution}

The ratios discussed above can be useful diagnostics of cloud evolutionary tracks.  They can describe the dynamic sources and sinks of the cloud as it proceeds through phases dominated by  H$_2$
formation, gas steady-state cycling where H$_2$ dissociation and formation are nearly in equilibrium and cloud dispersal.  Measuring these rates via H$_2$ emission could determine if clouds are in one phase of evolution or another. 

\begin{figure*}
    \centering
    \includegraphics[width=1\textwidth]{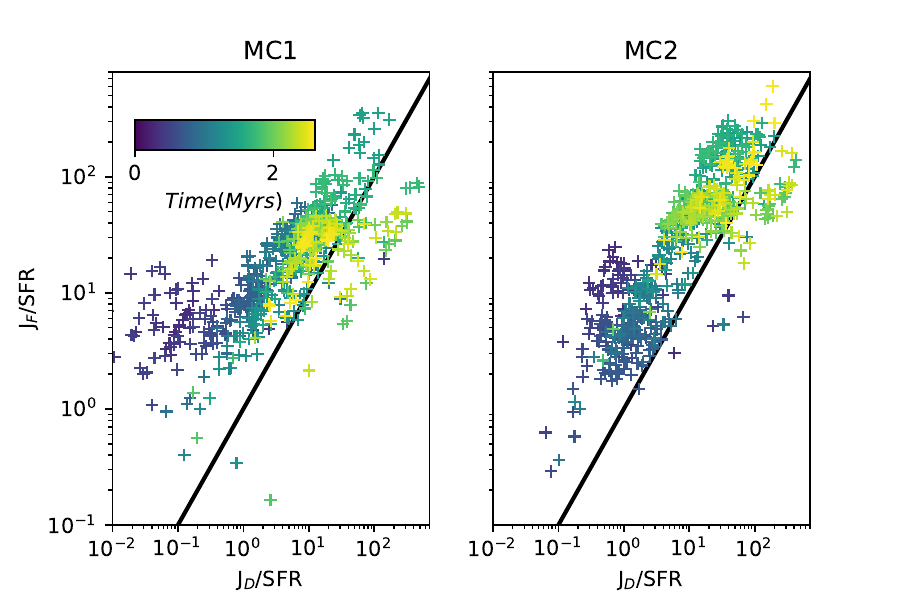}
     \caption{Cloud evolution for MC1 (left panel) and MC2 (right panel), comparing ratios of H$_2$ formation-to-SFR vs. H$_2$ dissociation-to-SFR. Points are from the numerical simulation and color indicates time as shown in the color bar inset in the figure. The diagonal black line is the 1-1 line, representing the H$_2$ formation-destruction steady-state ($J_F=J_D$).}
    \label{fig:ratio_sfr} 
\end{figure*}

To demonstrate this, we plot the ratios of $ J_F $/SFR vs. $ J_D $/SFR in Figure \ref{fig:ratio_sfr} for MC1 (left panel) and MC2 (right panel). The color of the points indicates the time evolution of the clouds,  as shown in the color bar inset in the figure.
At early times in the cloud's life (dark blue points), both MC1 and MC2 show that the H$_2$ formation rate dominates over other rates, and the cloud, therefore, builds its mass. In the SILCC-Zoom simulations, it has been demonstrated that this rapid molecular build-up is due to the compression of large-scale supernova remnant collisions \citep{2022arXiv220402511G},  which provides a sufficient shielding column of gas against the ISRF and UV from stars. The compression promotes both H$_2$ formation and then rapid star formation and the H$_2$ formation rate dominates over the H$_2$ dissociation rate and hence the points lie above the 1:1 line. 
MC2 has a larger H$_2$ dissociation rate early on than MC1 due to more rapid early star formation.  Hence it exits the cloud building stage sooner.

The early increase in the SFR also leads to a later increase in the H$_2$ dissociation rate via the increase in the stellar UV field. At this point, about 1.5 Myrs into the cloud lifetime, the cloud settles into a quasi-equilibrium phase where H$_2$ dissociation and formation are nearly balanced \citep{haid2019MNRAS.482.4062H}. From these Figures, it is  clear
that the SILCC-Zoom simulations at late times (yellow/green points) are nearly in an equilibrium scenario
where the MCs are in a bursty equilibrium between H$_2$
formation and destruction.

The overall picture with time evolution consists of cloud buildup, rapid star formation, followed by slower SFR, and near equilibrium between H$_2$ dissociation and formation. 
Eventually, the cloud will be fully dispersed by stellar feedback and supernova; however, these zoom-in runs were only followed for a little over 2.5 Myrs and in that time do not contain a supernova.  We further discuss Figure \ref{fig:ratio_sfr}'s connection with the lifecycle of molecular clouds and its \textit{observability} in the discussion section.

\section{Discussion}
\label{sec:dis}

\subsection{Cloud Evolution Scenarios: Disentangling the Life-cycle of H$_2$}
\begin{figure*}
    \centering
    \includegraphics[width=0.9\linewidth]{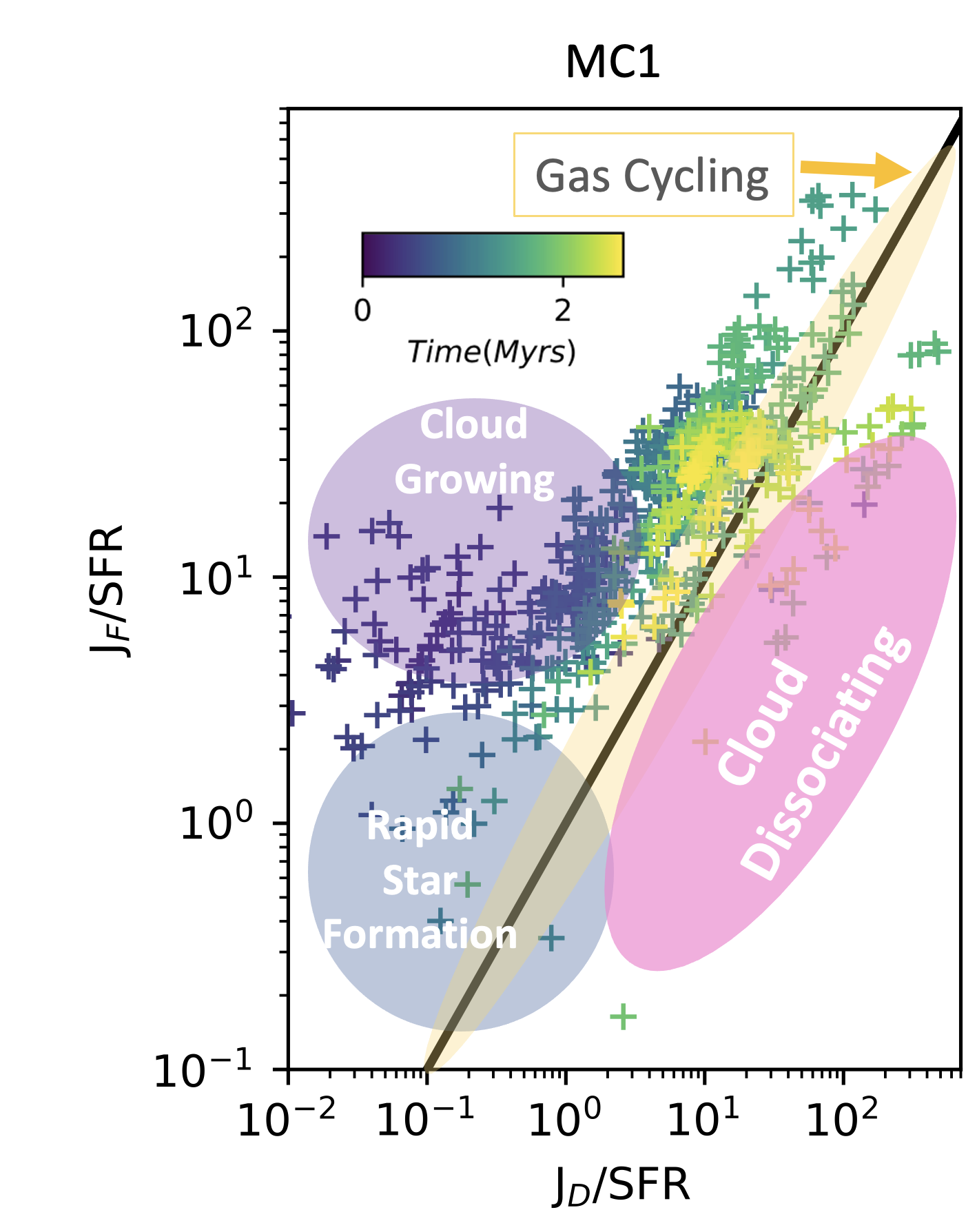}
     \caption{Cloud evolution for MC1, comparing ratios of H$_2$ formation-to-SFR vs. H$_2$ dissociation-to-SFR. Points are from the numerical simulation and color indicates time as shown in the color bar inset in Figure \ref{fig:ratio_sfr}. We overplot regions of parameter space in which different cloud evolutionary states exist, as discussed in Section~\ref{sec:dis}. The state of the cloud, as it proceeds through 
formation, gas steady-state cycling, and cloud dispersal, can be determined by the ensemble of
these rates. }
    \label{fig:cartoon} 
\end{figure*}

Given observations of molecular clouds only provide a snapshot of the cloud at one stage of its life, how could we measure what stage it is at provided data on the rates discussed above?  We add labeled regions of parameter space to Figure \ref{fig:ratio_sfr} MC1 and show the resulting cartoon in Figure \ref{fig:cartoon} for discussion here. This diagnostic plot demonstrates the ability to disentangle different cloud evolution scenarios if H$_2$ formation and dissociation rates can be measured (i.e.,  via H$_2$ fluorescent line emission, further discussed below). We overplot regions of parameter space in which different cloud evolutionary tracks exist. The state of the cloud, as it proceeds through a H$_2$  
formation dominant phase (purple area), enhanced SFR phase (blue area), gas steady-state cycling  phase(yellow), and cloud dispersal phase (pink area, bottom right corner), can be determined by the ensemble of
these rates.

In
a scenario where clouds have distinct evolutionary phases, they
are born in the upper left “cloud growing”
region (purple), and then rapidly form stars. 
The star formation increases the H$_2$ dissociation and the clouds enter a phase of near equilibrium which is indicated with labels with “H$_2$ equilibrium” region (yellow). Finally, clouds are dispersed fully by feedback or large scale shearing (if near the centers of galaxies) and would move to the lower right corner of the phase space.

How long does the steady-state scenario last before dissociation dominates?  Is dissociation dominated by mechanical dispersal (shocks, supernova) or by the UV field directly? Such questions could be answered if a statistical sampling of clouds in this parameter space is made.  In the extended
steady-state scenario, where \hh molecule lifetimes
are short,  observations of many molecular clouds would show most clouds distributed along the yellow region on the black one-to-one line, and concentrated
in the upper right in an extended steady
state  where \hh formation and destruction rates
greatly exceed star formation rates \citep{2021RNAAS...5..222K,2023arXiv230110251J}.

If MC formation
and dispersal is mainly mechanical and not chemical,
and molecules live much longer than clouds,
molecular clouds should concentrate in the lower left of Figure \ref{fig:cartoon}, “rapid
star formation”, where star formation rates exceed both \hh
formation and destruction. This sort of scenario is not observed in the two SILCC-Zoom simulation examples we explore here. This is likely because once star formation rapidly begins, stellar feedback immediately slows the formation of the next generation of stars and promotes additional H$_2$ dissociation via the production of ionizing photons.
Within Galactic molecular clouds,
different parts of the cloud may appear in different
regions on Figures \ref{fig:cartoon}.

We note that two sets of simulations are not enough to make statistical inferences which would ultimately be required to make a definitive statement about the underlying physics of MC dispersal and formation.  The simulations studied here do not capture changes in galactic environment, cloud surface density, or other important quantities such as metalicity, which could also affect cloud lifetimes. Future work will study the formation and destruction rates in an ensemble of clouds in different parts of the galaxy and in different galactic environments (Johnson et al. in prep).

\subsection{The Eos Space Telescope: Determining the Evolutionary Pathways of Star-Forming Clouds}

The H\,\textsc{I}/H$_2$ transition is complex, being affected by the heavy element content of the gas, the Galactic radiation field, and flows spanning spatial scales from spiral arms down to the thickness of a shock.
A recently proposed space telescope, Hyperion, is designed to probe star-forming clouds' evolutionary states systematically for the first time.
Hyperion carries an FUV spectrograph with resolution $R > 10,000$ covering wavelengths 138.5 to 161.5~nm \citep{2022JATIS...8d4008H}.
Hyperion was proposed to the 2021 Medium Explorer (MIDEX) call, and while receiving a Category I rating (the highest), was not selected.
A similar concept, Eos, is in development by the same team for the 2025 Small Explorer (SMEX) proposal call.

The primary scientific goal of Eos is to examine the fuel for star formation by probing the crucial atomic-to-molecular boundary layer in and around molecular clouds.
The telescope must observe from space since the FUV band is absorbed by the Earth's atmosphere.
Measuring the FUV fluorescence spectrum of H$_2$ will enable Eos to determine conditions at molecular clouds' surface, where they interact with the surrounding Galactic environment.
The fluxes into and out of the clouds govern their evolutionary history, present state, and possible futures, as shown by Figure~\ref{fig:cartoon}.
Because the H$_2$ UV fluorescent lines are intrinsically narrow, numerous, and closely-spaced in wavelength for a range of excitation conditions, high-resolution spectroscopy is essential for this mission.
The spectral resolution of Eos will also permit detecting flows in the fluorescing gas down to the 10~km s$^{-1}$ range.
Measuring these parameters will uncover the relationship in local Milky Way molecular clouds between cloud growth, star formation, cloud evaporation in the presence of a radiation field, and young stars' impacts on subsequent star formation.

\subsection{Observations to Measure the Rates Described in this Work}

Measuring the dissociation rate $J_D$ is straightforward with Eos, since the H$_2$ photodissociation rate is proportional to the H$_2$ photo-excitation rate, which in turn is proportional to the intensity of H$_2$ line emission 
\citep{1989Sternberg}.

Observationally, probing H$_2$ formation rates with FUV lines is not as technically simple as measuring dissociation. A key signature of the formation of H$_2$ found in the FUV fluorescence emission lines is the ortho-to-para line ratio (OPR).  
In the ground state, ortho-H$_2$ molecules exhibit aligned spins (total nuclear spin = 1, only odd rotational quantum numbers), while para-H$_2$ molecules possess spins in opposite directions. These distinct states result in subtly different permitted transitions for fluorescence.
Under equilibrium conditions within a photodissociation region (PDR), the OPR in the ground vibrational state is determined by the competition between reactive collisions with protons, the H$_2$ formation process, and selective photodissociation via optically thick ("self-shielded") absorption lines \citep{Sternberg_1999}. A low OPR provides unambiguous evidence of older, colder H$_2$ gas, as there is no ambiguity in the gas state. On the other hand, a high OPR can arise from two different processes: One possibility is a warm PDR that has already reached chemical equilibrium, while the other is cold, young H$_2$ gas that has not yet attained chemical equilibrium and possesses a high OPR due to the formation process on dust grains. To differentiate between the two scenarios, temperature measurements can be self-consistently estimated using H$_2$ fluorescent emission. In the process of mapping target clouds, Eos will collect absorption line measurements, which can provide insights into cold H$_2$ within certain parts of the cloud. In these regions, the equilibration time is unambiguously long, allowing for a straightforward interpretation of the OPR measurement as a chemical age indicator. Paper I (Bialy et al. in prep)  further describes observational probes of the formation and dissociation rates and tests the assumptions of equilibrium HI-H$_2$ time scales.

\section{Conclusions}
\label{sec:con}
In this paper we used the SImulating the LifeCycle of molecular Clouds (SILCC)-Zoom simulations that include stellar feedback to explore the
utility of FUV H$_2$ fluorescence observations to trace the life-cycle of star-forming molecular clouds and
the molecular hydrogen to atomic hydrogen boundary layer.  The simulations used in this work include non-equilibrium H$_2$ formation/destruction, stellar radiation, sink particles, and turbulence.  We compute the star formation rate, H$_2$ dissociation rate and formation rate and find that:
\begin{itemize}
    \item At early times in the cloud evolution,  H$_2$ formation rapidly outpaces dissociation and molecular clouds build their mass. 
    \item We find that the H$_2$ is  out of chemical equilibrium during the early stages of cloud formation and rapid star formation. 
    \item The concept of an integrated star formation efficiency is not applicable to clouds with realistic open boundaries, as H$_2$ formation and cloud building continue after star formation is initiated. 
    \item H$_2$ formation/dissociation equilibrium  takes place  after the cloud is formed but before it is dispersed by massive star feedback.
    \item Cloud life-cycle scenarios and star formation efficiency theories may be distinguished with observational measurements of H$_2$ fluorescence with a high-resolution FUV spectrometer covering large angular areas of nearby molecular clouds, such as the proposed Hyperion and Eos missions. 

\end{itemize}

 \acknowledgements

 B.B. acknowledges support from NSF grant AST-2009679 and NASA grant No. 80NSSC20K0500.
B.B. is grateful for generous support by the David and Lucile Packard Foundation and Alfred P. Sloan Foundation. 
DS and SW acknowledge funding support from the Deutsche Forschungsgemeinschaft (DFG) via the
Sonderforschungsbereich (SFB) 956, \textit{Conditions and Impact of Star Formation} (projects C5 and C6). Furthermore, the project is supported by the B3D project, a program within ``Profilbildung 2020", an initiative of the Ministry of Culture and Science of the State of Northrhine Westphalia. The sole responsibility for the content of this publication lies with the authors.
TJH is funded by a Royal Society Dorothy Hodgkin Fellowship and UKRI guaranteed funding for a Horizon Europe ERC consolidator grant (EP/Y024710/1). A.S. and B.B. thank the Center for Computational Astrophysics (CCA) of the Flatiron Institute and the Mathematics and Physical Sciences (MPS) division of the Simons Foundation for support. MRK acknowledges support from the Australian Research Council through the Laureate Fellowship program, award FL220100020.  The research was carried out in part at the Jet Propulsion Laboratory, California Institute of Technology, under contract 80NM0018D0004 with the National Aeronautics and Space Administration.
\bibliographystyle{apj}

\bibliography{ms}

\end{document}